\documentclass[fleqn,usenatbib]{mnras}

\usepackage{newtxtext,newtxmath}

\usepackage[T1]{fontenc}

\DeclareRobustCommand{\VAN}[3]{#2}
\let\VANthebibliography\thebibliography
\def\thebibliography{\DeclareRobustCommand{\VAN}[3]{##3}\VANthebibliography}

\usepackage{graphicx}	
\usepackage{amsmath}	
\usepackage{tabularx}

\title{Variations in the Radiation Intensity of Pulsar B0950+08:
	Nine Years of Monitoring at 110 MHz}

\author[T. V. Smirnova et al.]{
T. V. Smirnova $^{1}$\thanks{E-mail: tania@prao.ru}
M. O. Toropov,$^{2}$
S. A. Tyul'bashev,$^{1}$
\\
$^{1}$ Pushchino Radio Astronomy Observatory, Astro Space Center, Lebedev Physical Institute, Russian Academy of Sciences, Pushchino, 142290 Russia \\
$^{2}$ LLC TEK Inform, Moscow, 117246 Russia \\
}

\date{2024}

\pubyear{2024}

\begin{document}
\label{firstpage}
\pagerange{\pageref{firstpage}--\pageref{lastpage}}
\maketitle

\begin{abstract}
The analysis of variations in the emission intensity of the pulsar B0950+08 from 2014 to 2022 with scales from minutes to years was carried out. The observations were obtained in a round-the-clock daily survey conducted on the Large Phased Array (LPA) radio telescope. The high variability of emission is shown not only from pulse to pulse, but also at scales greater than 3 min. The average value of the estimated amplitude of these variations in 3.2 minutes is 25~Jy, the modulation index is 1. The average relative amplitude of the interpulse (IP) is $2.00 \pm 0.28\%$ of the main pulse. In individual pulses, the amplitude of the interpulse may exceed the amplitude of the main pulse (MP), but this is a rare event. Emission is observed in almost the entire period of the pulsar. For the first time, the relative amplitude of emission between the main pulse and the interpulse (emission bridge) was measured. When averaging about 10 hours, it varies from 0.8\% to 1.31\% with an average value of $1.04 \pm 0.28\%$.

A high correlation was found between MP and IP amplitude variations both when averaging profiles over 3.2 minutes and when averaging over years. This correlation is due to refractive interstellar scintillation. The frequency scale of IP diffraction interstellar scintillation was measured for the first time and it was shown that the spectral forms for IP and MP are well correlated and have the same frequency scale. There are strong variations in the frequency scale of scintillation $f_{dif}$ from session to session (time interval from one day) on scales of 200-800 kHz. The refractive scale of scintillation for 1-2 days has been determined. A modulation of emission with a characteristic scale of about 130 days was detected, which, apparently, is also associated with refractive scintillation.
\end{abstract}




\section{Introduction}

Pulsar B0950+08 (J0953+0755) is one of the most powerful pulsars in the meter range. It has low-level emission for almost the entire period of (\citeauthor{Hankins1981}, \citeyear{Hankins1981}; \citeauthor{Perry1985}, \citeyear{Perry1985}; \citeauthor{Smirnova1988}, \citeyear{Smirnova1988}). Apparently, this is due to the fact that the angle between the magnetic axis and the axis of rotation is small (the case of a coaxial rotator) and the observer sees the emission area during the entire period (\citeauthor{Lyne1988}, \citeyear{Lyne1988}). In addition to the main pulse (MP), the pulsar has an interpulse (IP), a prepulse and a emission bridge (M) between MP and IP. The pulsar is characterized by a high degree of polarization: $67 \pm 6\%$ at a frequency of 39 MHz (\citeauthor{Suleimanova1983}, \citeyear{Suleimanova1983}), and the positional angle smoothly changes by 180 degrees from MP to IP (\citeauthor{Backer1980}, \citeyear{Backer1980}), located at a distance of 152 degrees from MP. A number of studies indicate a high variability in the intensity of individual pulses and the presence of giant pulses for this pulsar (\citeauthor{Smirnova2006}, \citeyear{Smirnova2006}; \citeauthor{Smirnova2012}, \citeyear{Smirnova2012}; \citeauthor{Singal2012}, \citeyear{Singal2012}; \citeauthor{Kuiack2020}, \citeyear{Kuiack2020}). Although a lot of observational data has been accumulated for the pulsar in a wide frequency range, however, many questions remain about the geometry of its magnetosphere, the localization of the radio emission region and the mechanism of emission.

\begin{figure*}
\begin{center}
	\includegraphics[width=0.7\textwidth]{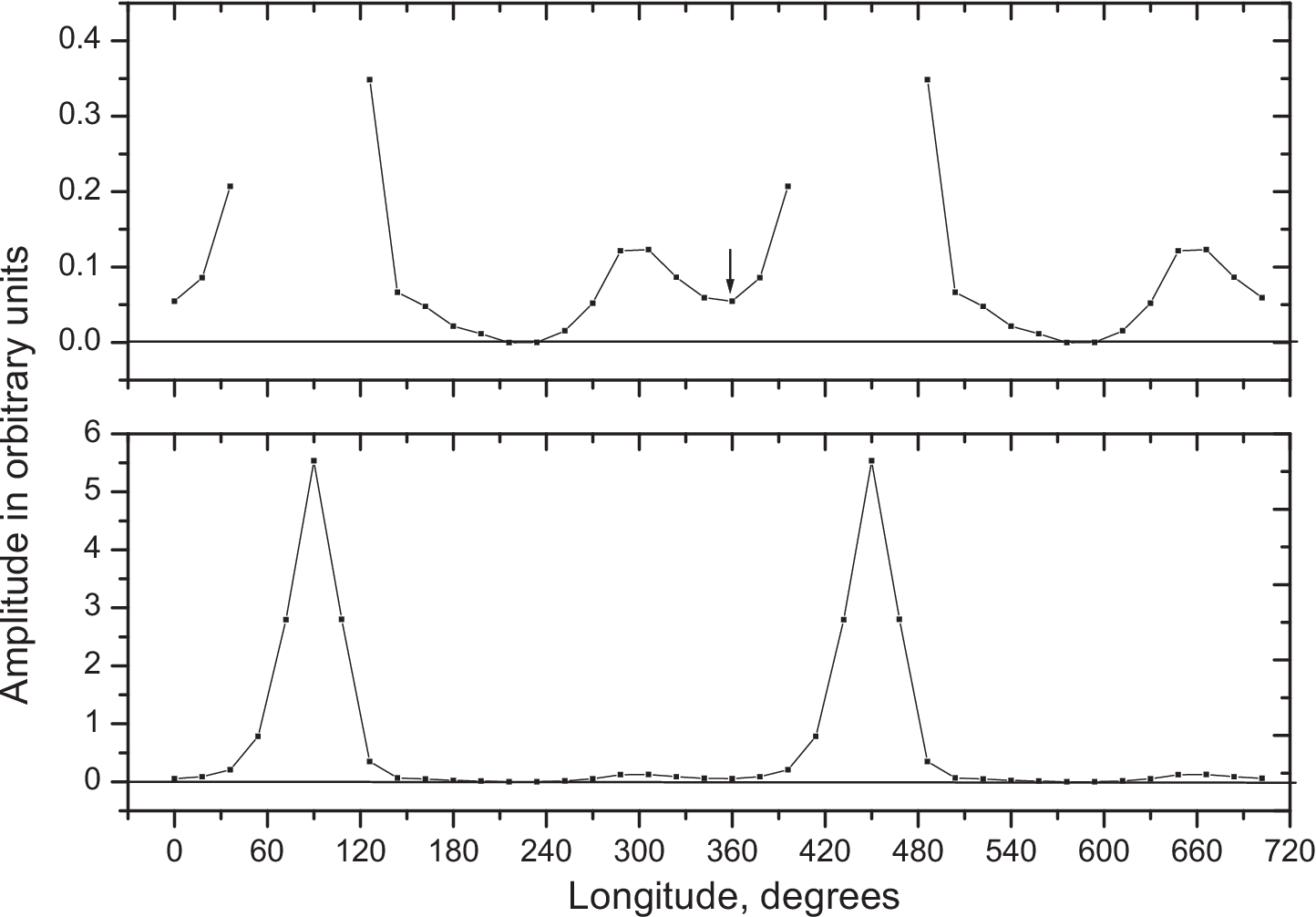}
    \caption{The average profile is B0950+08, accumulated in 2014 (112 sessions). The arrow marks the longitude at which the amplitude of the bridge (M) was measured. This profile is shown at the top in an enlarged scale so that the inter-pulse emission is better visible.}
    \label{fig:fig1}
\end{center}
\end{figure*}

The passage of emission from pulsars through the inhomogeneities of interstellar plasma leads to a number of observed effects: angular broadening of the source, temporal broadening of pulses, modulation of the emission intensity in frequency and time (scintillating). The study of these effects makes it possible to investigate the spatial structure of the inhomogeneities of interstellar plasma. During the propagation of emission, radio waves are scattered on inhomogeneities of interstellar plasma into a certain angular spectrum with a scale of $\theta_{scat}$, called the scattering angle. At some distance from the scattering layer r, their interference leads to amplitude modulation. Diffraction scintillation are distinguished at the scales of $s_{dif} = \lambda/\theta_{scat}$ ($s_{dif} \sim 10^7$~cm by $\lambda = 1$~m) and refractive scintillation, which are realized on significantly large spatial scales: $s_{ref} = \theta_{scat}r$ ($s_{ref}\sim 10^{12}$~cm, $r$ is the distance from the observer to the scattering screen). A large number of experimental facts have been interpreted within the framework of homogeneous isotropic Kolmogorov turbulence \citeauthor{Armstrong1995} (\citeyear{Armstrong1995}); \citeauthor{Shishov2002} (\citeyear{Shishov2002}). However, it was shown that in some directions the exponent of the degree of the spectrum of inhomogeneities, $n$, differs significantly from Kolmogorov's: $n = 11/3$. In particular, in the direction of the pulsar B0950+08 $n = 3$ (\citeauthor{Smirnova2008}, \citeyear{Smirnova2008}; \citeauthor{Smirnova2014}, \citeyear{Smirnova2014}). According to observations on the Radioastron ground-space interferometer, it was also shown \citep{Smirnova2014} that in the direction B0950+08 there are 2 scattering layers at distances of 4.4-16.4 pc and 26-170 pc. In the work (\citeauthor{Smirnova2008}, \citeyear{Smirnova2008}), characteristic scintillation scales were obtained for this pulsar at four frequencies 41, 62, 88 and 110 MHz.

In this paper, over an observation interval of 9 years, we investigate both the behavior of the main components of the pulsed emission of the pulsar B0950+08 and the influence of inhomogeneities of interstellar plasma on the passage of its emission to the observer.

\begin{figure*}
\begin{center}
	\includegraphics[width=0.7\textwidth]{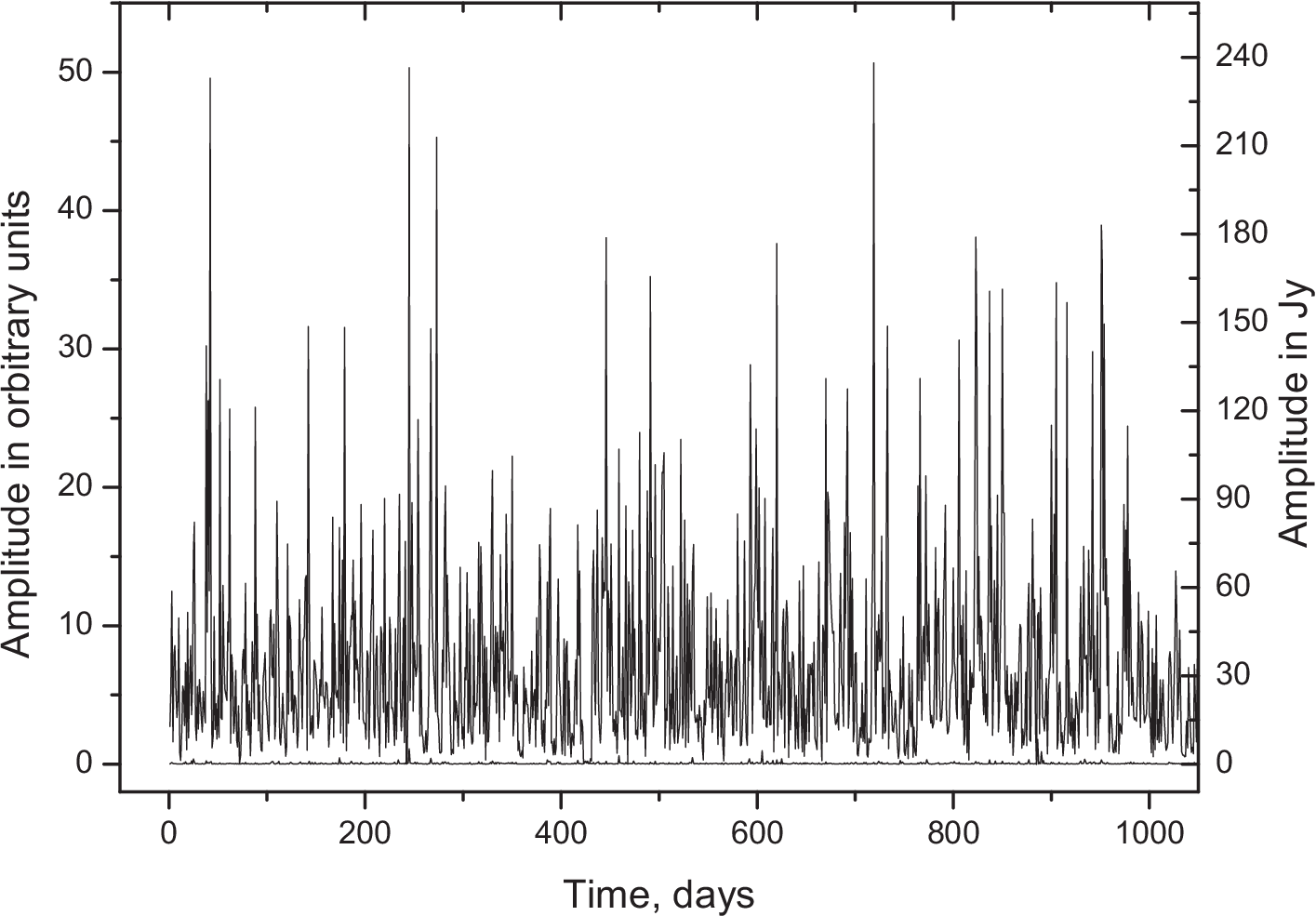}
    \caption{The dependence of the MP amplitude for the average profile per session on time. At the bottom, $\sigma_n$ from time is shown at the zero level. The y axis on the left is the amplitude in relative units, on the right is in Jy. The first day corresponds to 08/21/2014 (MJD 56890).}
    \label{fig:fig2}
\end{center}
\end{figure*}

\section{Observations and primary data processing}

The observations were carried out on the Large Phased Array (LPA3) of the Lebedev Physical Institute (LPI) under the Pushchino Multibeams Pulsar Search program (PUMPS) (\citeauthor{Tyulbashev2016}, \citeyear{Tyulbashev2016}; \citeauthor{Tyulbashev2022}, \citeyear{Tyulbashev2022}). During the reconstruction of the LPA, two independent radio telescopes were created on the basis of one antenna field. One of them (LPA3) has 128 fixed beams overlapping declinations of $-9^{\circ} < \delta < +55^{\circ}$ at the intersection level of 0.4. The LPA is built on dipoles that receive linearly polarized emission in one plane.

Pulsar B0950+08 enters the observation area and has been observed daily for 9 years. The central frequency of observations is 110.3 MHz, the reception band is 2.5 MHz. Observations are carried out on the antenna in different modes and on different receivers. In this work, a 32-channel receiver with a channel width of 78 kHz was used. The sampling
 time of the point is 12.5 ms. To calibrate the signal in the frequency channels, a calibration step with a known temperature was used, which was recorded 6 times a day. Daily pulsar recordings were conducted from 08/21/2014 to 12/31/2022 (MJD 56890-59944), for a total of 3054 days.

The data was recorded in all frequency channels on the disk in hourly portions. The part corresponding to the time of passage of the pulsar through the antenna  at the half-power level was selected from the corresponding hour record. For the pulsar B0950+08, this time is 3.2 minutes (798 pulses). The primary processing included several stages: calibration of the signal according to the calibration step so that the gain in all channels was the same; subtraction of the baseline; compensation of dispersion; recording of all pulses in all frequency channels on a disk. After that, the average profiles obtained for each day were analyzed, and sessions in which the quality of the average profile was low were rejected.

The average profile for each session was obtained after compensation of the variance by adding all records with a given period. The average profile was cyclically shifted so that MP was in the first quarter of the period. The MP phase was the same in all sessions, so you could add them up to get an average profile for all sessions of a given year. Fig.~\ref{fig:fig1} shows one of these profiles, accumulated over 112 sessions in 2014. For greater clarity, it is given with a double period. The components of this profile are clearly visible here: MP, IP and bridge. Emission takes place almost the entire period.

The pulsar was observed in two adjacent beams, so it was possible to monitor the state of the ionosphere and if the offset of the declination coordinate was large, then these sessions were not used for analysis. In the presence of strong interference, such recordings were also excluded from the analysis. In addition, some of the days were used for technical work on the antenna. In total, 9.7\% of the total number of sessions were excluded for the reasons listed above. Since the pulsar emission takes up almost the entire period, for each session, a time series of amplitudes (positive and negative deviations from the average level) was formed at the longitude of the minimum value of the amplitude of the average profile, and then the value of the noise sigma ($\sigma_n$) was found along them. For each session, the values of $\sigma_n$, MP amplitude, IP and bridge for the average profile were recorded in a separate table for all observation days. The longitude at which the amplitude of the bridge was determined is marked with an arrow in Fig.~\ref{fig:fig1}.

Pulsar period $P$=0.253~s, dispersion measure $DM$=2.97~pc/cm$^3$, rotation measure $RM$=1.35~rad/m$^2$ according to the ATNF catalog (https://www.atnf.csiro.au/research/pulsar/psrcat /; (\citeauthor{Manchester2005}, \citeyear{Manchester2005}). Dispersion smearing in the band of one channel is less than a secret. In the average pulsar profile, we have 20 points with a resolution of 12.5 ms. The width of the average profile at the level of half the amplitude at a frequency of 111 MHz is 15 ms, the distance between the components is 6.2 ms (\citeauthor{Smirnova2012}, \citeyear{Smirnova2012}), therefore, the details of individual pulses (components and pre-pulse) are not resolved in our observations. In this work, we will analyze only three components: MP, IP and bridge.

\begin{figure*}
\begin{center}
	\includegraphics[width=0.7\textwidth]{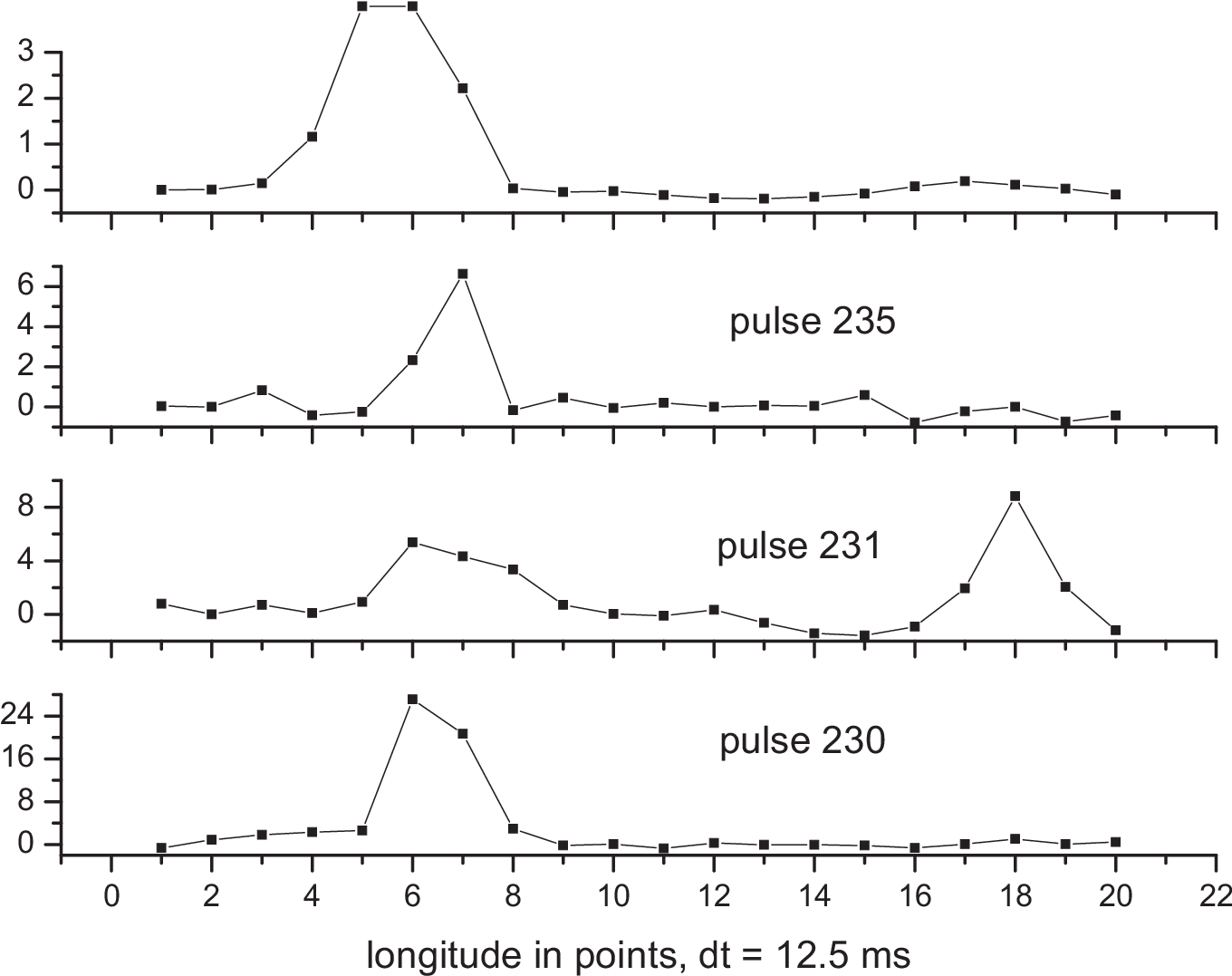}
    \caption{Three individual pulses (bottom -up) and the average profile for this session. The top of the middle profile is cut off (its amplitude is 10) so that the IP is better visible. The y-axis is the amplitude in relative units, the x-axis is the longitude in points.}
    \label{fig:fig3}
\end{center}
\end{figure*}

\section{Analysis and results}

Pulsar 0950+08 is one of the most powerful and closest pulsars. The distance to it is $R = 262 \pm 5$~pc, the transverse speed of the visual beam is $V_{perp} = 36.6\pm 0.7$ km/s (\citeauthor{Brisken2002}, \citeyear{Brisken2002}). In the work of Malofeev et al. (\citeauthor{Malofeev2000}, \citeyear{Malofeev2000}), the flux density was measured at a frequency of $f = 102.5$~MHz: $S = 2$~Jy. Figure 2 shows variations in the MP amplitude for the average profile per session (3.2 min) over the first 1000 days of observations. It can be seen that the amplitude varies greatly from day to day. The modulation index measured at MP longitude is $m = 0.9851$. It was determined by the formula:

\begin{equation}
	m = \sum_{i=1}^N \left[\frac{(A-\langle{A}\rangle )^2}{N-1}\right]^{1/2}/\langle{I}\rangle,
	\label{eq:1}
\end{equation}

where $A$ and $\langle{A}\rangle$ are intensity and average intensity, and $N$ is the number of points.
Strong variations are caused by the effects of emission propagation in inhomogeneous interstellar plasma. These effects include: Faraday rotation of the polarization plane, diffraction and refractive scintillation. Characteristic time and frequency scales of scintillating at 112 MHz: $t_{dif} > 200$~s and $f_{dif} = 220$~kHz (\citeauthor{Smirnova2008}, \citeyear{Smirnova2008}). Since the scintillation time scale is longer than the time of one session, and the frequency scale is smaller than the receiver band (2.5 MHz), the main contribution to the amplitude variations of the average profile per session is determined by the scintillation time scale.

The pulsar has a strong linear polarization, $P_l$=70-80\% at a frequency of 111 MHz (\citeauthor{Shabanova2004}, \citeyear{Shabanova2004}) and therefore polarization can also affect variations in the signal amplitude. The Faraday rotation period $P_F$[kHz]=$\pi\times f^3\times 10^5/18RM$, where $f$ is the frequency of observations in hundreds of MHz. Accordingly, $P_F = 17.3 $~ MHz at a frequency of 110.3 MHz, which is significantly more than the receiver band. Rotation of the polarization plane in the receiver band leads to a relative change in the amplitude of the profile components day by day. Since the positional angle between MP and IP varies by about 180 degrees (\citeauthor{Backer1980}, \citeyear{Backer1980}), the ratio of MP and IP amplitudes will not change from session to session, and the ratio of emission intensities in MP and bridge will be due to the difference in positional angles at these longitudes.

\section{Variations in the amplitudes of the components of the average profile}

Variations in the amplitude of $A(t)$ for MP, determined by the average profiles per session, are shown for the first 1050 days in Fig.~\ref{fig:fig2}. As can be seen from this figure, the amplitude varies significantly from day to day (more than 500 times). This is due to both polarization and diffraction and refractive scintillating during the propagation of emission through inhomogeneities of interstellar plasma. To bind the amplitude in relative units of $A(t)$ to the amplitude in Jy, we used the following relation:

\begin{equation}
A(t) =\frac{ A(t) \times S \times k_1}{\langle A \rangle}, 
	\label{eq:2}             
\end{equation}

where S = 2 Jy is the pulsar flux density at 111 MHz (\citeauthor{Malofeev2000}, \citeyear{Malofeev2000}), $k_1 = 12.17$ is a coefficient that takes into account the ratio of peak amplitude to pulse energy averaged over the pulsar period, $\langle A\rangle$ is the average value of the amplitude in relative units over the entire observation period. The energy in the pulse was calculated as the sum of the intensities within the MP boundaries for the average profile multiplied by the time step between the points (12.5 ms). Then the amplitude value is: $A(t)$~[Jy] $= A(t) \times 4.7$ and its average over the observation period is 34.1 Jy, and the peak value reaches 240 Jy. In Fig.~\ref{fig:fig2} the y axis on the right is given in Jy. The average value of the noise sigma $\sigma_n = 0.32$ in relative units. The largest amplitudes have a signal-to-noise ratio of $S/N = 750$.

To get the average amplitudes of MP, IP and bridge over the years, as well as their ratios, we averaged all profiles for each year. The values of the average amplitudes of MP, IP, and bridge ($A_{MP}$, $A_{IP}$, $A_M$) and their ratios by year are shown in Table~\ref{tab:tab1}. With this averaging, all the effects of emission propagation in interstellar plasma and polarization no longer have an effect. Averaging corresponds to an accumulation of about 16 hours (300 sessions). The average values of the amplitude of the profile components in Jy are given at the end of the table. The relative amplitude of the interpulse is $2.0 \pm 0.28\%$, however, in some sessions (with an accumulation of 3.2 minutes) it can reach up to 5\%. Pulses may be observed within a single session in which the amplitude of the interpulse may exceed the amplitude at the MP longitude. Such an example is shown in Fig.~\ref{fig:fig3}.

\begin{table*}
\centering
\caption{The amplitude ratios of the components of the B0950+08 profile and their peak flow densities by year.}
\label{tab:tab1}
\begin{tabular}{c|c|c|c|c|c}
\hline
year & $A_{IP}/A_{MP}$,\% & $A_M/A_{MP}$,\% & $A_{MP}$, Jy & $A_M$, Jy & $A_{IP}$, Jy\\
\hline
2014	&	2.2	&	1.01	&	25.77	&	0.26	&	0.57	\\
2015	&	2.15	&	1.31	&	24.27	&	0.31	&	0.49	\\
2016	&	1.93	&	1.37	&	25.95	&	0.25	&	0.50	\\
2017	&	1.9	&	0.78	&	24.38	&	0.19	&	0.46	\\
2018	&	2.3	&	1.15	&	22.56	&	0.26	&	0.43	\\
2019	&	2.16	&	1.15	&	24.56	&	0.29	&	0.54	\\
2020	&	2.14	&	1.14	&	21.20	&	0.24	&	0.45	\\
2021	&	1.8	&	0.98	&	24.38	&	0.22	&	0.44	\\
2022	&	1.4	&	0.47	&	20.49	&	0.10	&	0.29	\\
average & $2.0 \pm 0.28$ & $1.04 \pm 0.28$ & $23.7 \pm 1.9$	& $0.24 \pm 0.06$ & $0.46 \pm 0.08$\\
\hline
\end{tabular}
\label{tab:tab1}
\end{table*}

\begin{figure*}
\begin{center}
    \includegraphics[width=0.5\textwidth]{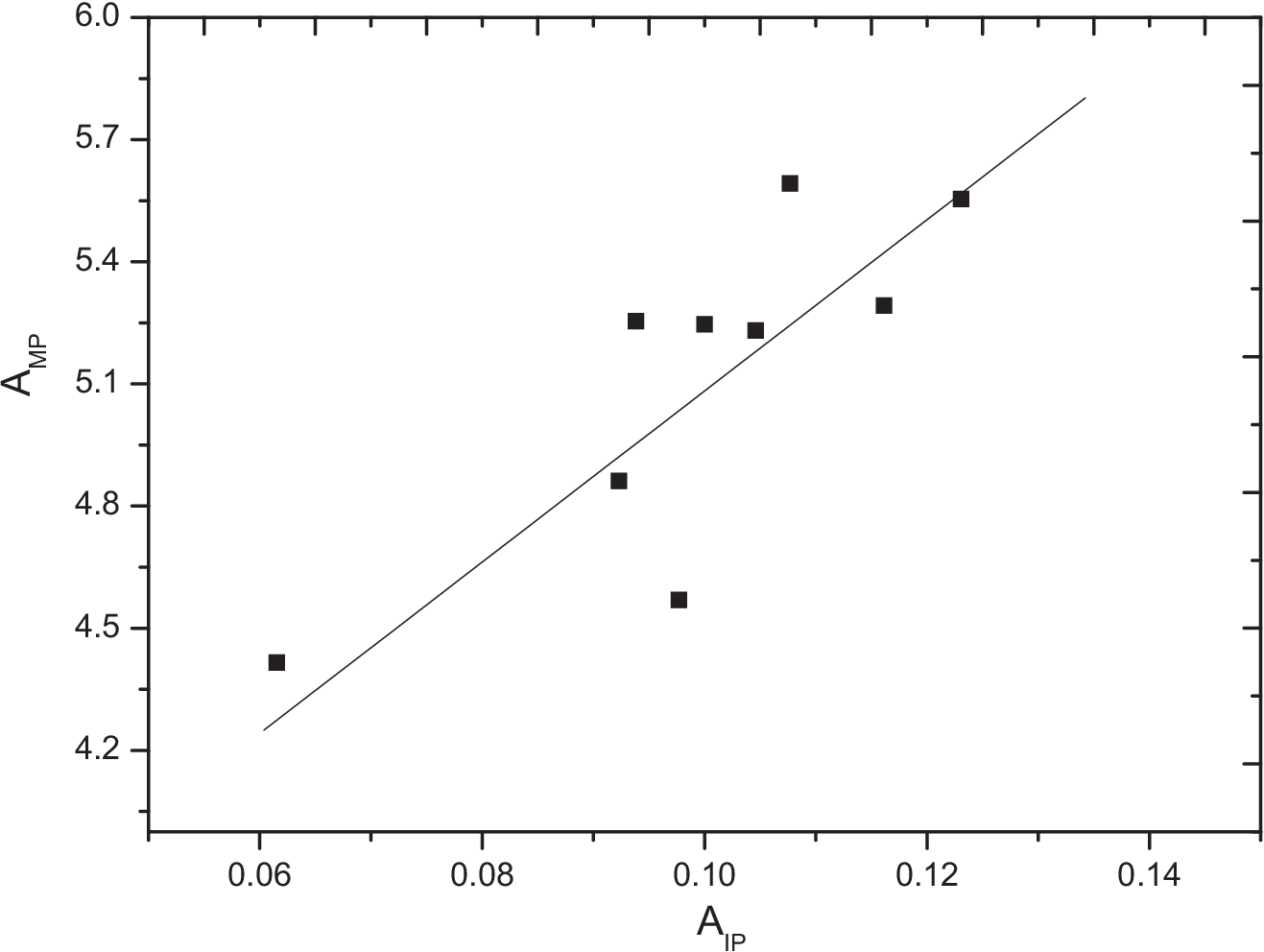}
    \caption{Average profiles accumulated over 10 years (3101 sessions) J2234+2114. The x-axis is the time in ms, the y-axis is the amplitude of the profile in relative units.}
    \label{fig:fig4}
\end{center}
\end{figure*}

Here are 3 individual pulses and the average profile for one of the sessions. In pulse number 231, the amplitude of the interpulse exceeds the amplitude of MP by 1.64 times. Fig.~\ref{fig:fig4} shows the dependencies between the average values of the amplitudes of the profile components over 9 years of observations. As can be seen from this figure, there is a correlation between the amplitudes of MP and IP when averaged over the years. The correlation coefficient for all points is $R = 0.75$, if you remove the value for 2022 (the point in the lower left corner), then $R = 0.64$. There is no correlation for the other components.

\begin{figure*}
\begin{center}
	\includegraphics[width=0.7\textwidth]{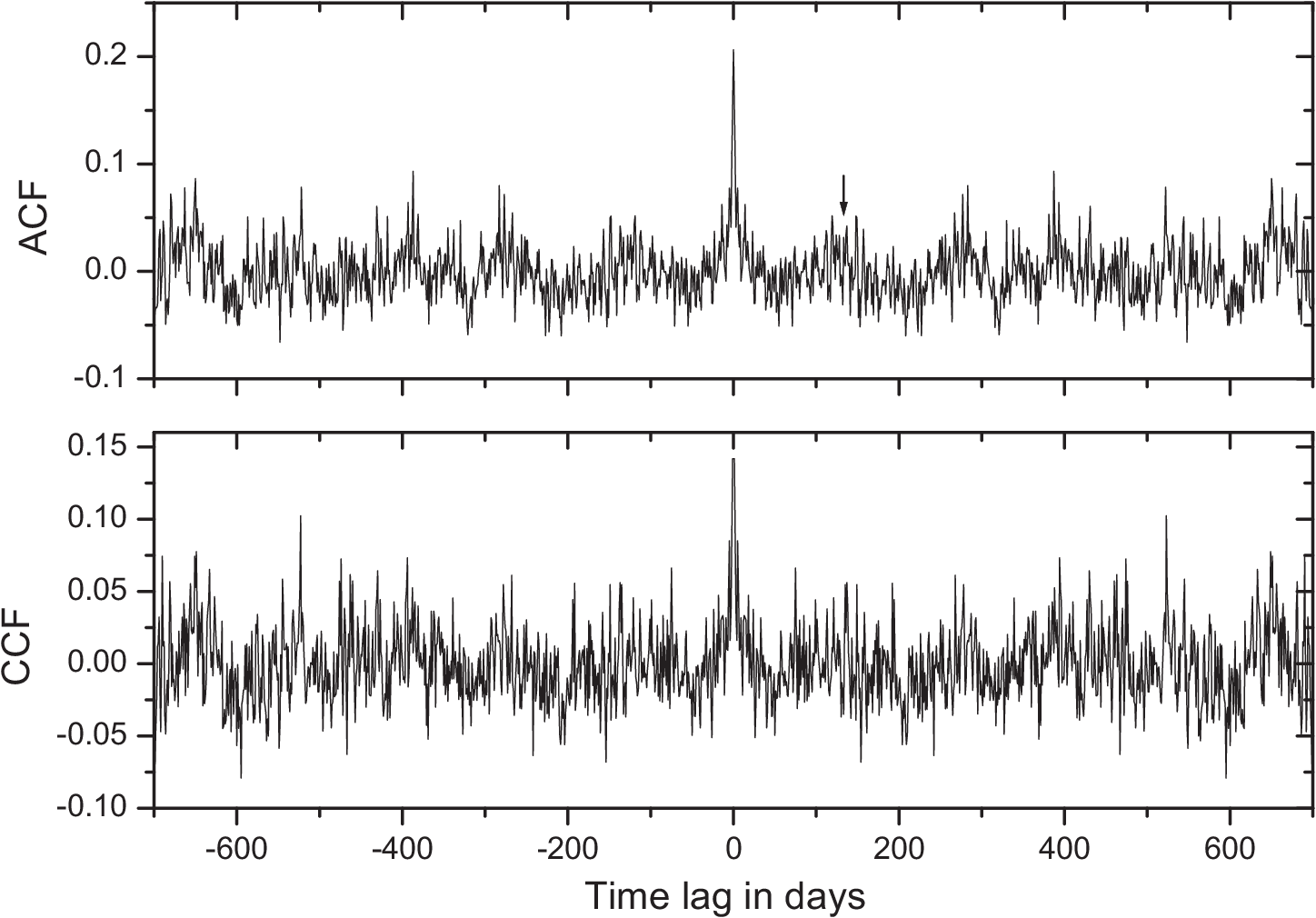}
    \caption{Cross-correlation function between the amplitudes of the average MP and IP profiles per session (lower panel), at zero shift, the CCF value (0.63 for $k=0$) is replaced by the value at $k = 1$; autocorrelation function from variations in MP amplitudes (upper panel). The arrow marks a shift of 130 days.}
    \label{fig:fig5}
\end{center}
\end{figure*}

\section{Correlation analysis}

To see if the amplitude variations of MP and IP obtained from the average profiles per observation session (3.2 min) correlate, we calculated the cross-correlation function (CCF) between them over the entire observation interval. Since we have uneven (with missing bad sessions) rows, we defined CCF as:

\begin{equation}
\begin{array}{cc}
     CCF(k) = \frac{1}{(\sigma_1 \times \sigma_2 \times N(k))} \times & \\
     \times \sum^{M-k}_{i=1}  g( i) g(i + k)[(I_1 (i) - \langle I_1 \rangle ) \times (I_2 (i+k) - \langle I_2 \rangle) ], & \\
\end{array}
	\label{eq:3}
\end{equation}

where $\langle I_1\rangle$ and $\langle I_2 \rangle$ are the average values of MP and IP amplitudes, and $\sigma_1$ and $\sigma_2$ are standard deviations, $M$ is the length of the array, $k$ is the time shift in days, $k = 0, 1, 2, ..., 0.8M$, $g(i) = 1$ when there is a value on a given day i and is 0 otherwise. $N(k) = \sum g(i)g(i+k)$ is the number of points at a given shift of $k$. The autocorrelation function (ACF) is calculated in the same way when replacing $I_2$ with $I_1$. The calculation starts with $k = 1$ to eliminate the effect of noise at zero shift for ACF. For CCF, we did not count the function for negative shifts, but simply mirrored the function for negative shifts. Fig.~\ref{fig:fig5} shows the CCF between MP and IP amplitudes, as well as the autocorrelation function of MP amplitude variations. CCF ($k = 0$) is 0.63. This value is not shown in the figure to better see the smaller CCF values. Obviously, there is a high correlation of MP and IP amplitudes for the same sessions. Noise at MP and IP longitudes is not correlated at zero shift. These correlated variations of MP and IP are due to refraction on the same medium and are not intrinsic variations of pulsar emission. With a 1-day shift, CCF ($k = 1$) is 0.14, and ACF ($k = 1$) = 0.21. Even with a 2-day shift, there is a significant correlation: CCF ($k = 2$) is 0.09, and ACF ($k = 2$) is 0.11. If we determine the characteristic decorrelation time at a shift at which the ACF drops by a factor of 2, then this is $T_{ref} = 2$~days. In addition to this 1-2-day scale, there seems to be a slow modulation with a scale of about 130 days, marked with an arrow in Fig.~\ref{fig:fig5}. The correlation of these variations is due to refraction on larger scales.

\begin{figure*}
\begin{center}
	\includegraphics[width=0.6\textwidth]{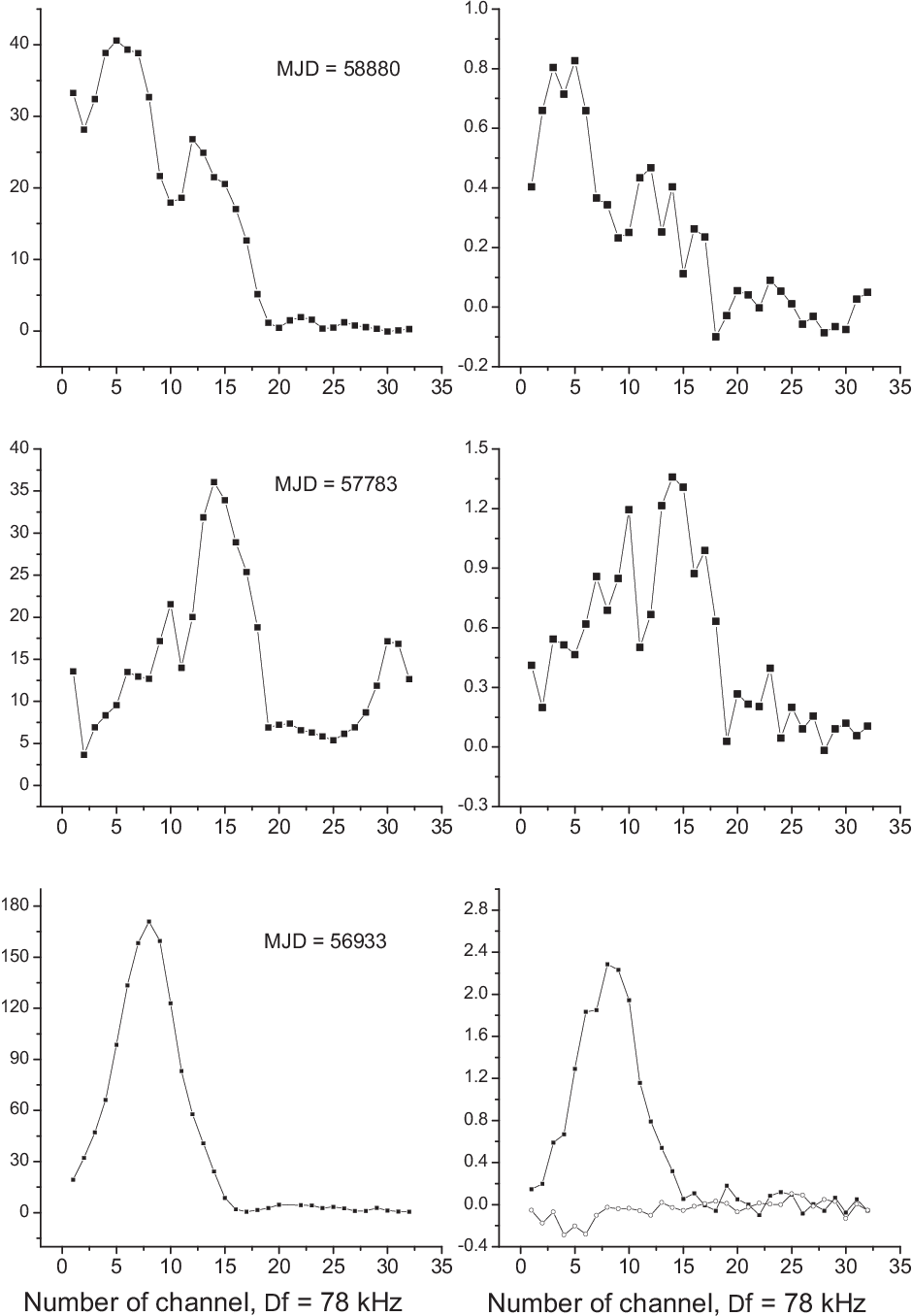} 
    \caption{The average spectra for 3 observation sessions at MP (left) and IP (right) longitudes. The amplitudes are given in relative units. For MJD = 56933, the noise spectrum is additionally shown on the IP spectrum.}
    \label{fig:fig6}
\end{center}
\end{figure*}

\section{The influence of interstellar plasma}

When radiation propagates through inhomogeneities of interstellar plasma, the effects of scintillating are most pronounced at low frequencies. On meter waves, the scintillation are saturated and both diffraction and refractive scintillation are realized. Characteristic time scales of diffraction scintillating, $t_{dif} = s_{dif} / V_{ef}$, and refractive scintillation $T_{ref} = s_{ref} / V_{ef}$. At a frequency of 110 MHz, for diffraction scintillation, these are seconds and minutes, and for refractive scintillation, these can be days, months and years. Here $V_{ef}$ is the velocity of the diffraction pattern relative to the observer. This speed consists of three components:

\begin{equation}
V_{ef} = r V_{psr} / (R-r) + {\bf V}_{obs} - R / (R-r) V_{scr},
	\label{eq:4}                  
\end{equation}

where $V_{psr}$, ${\bf V}_{obs}$ and $V_{scr}$ are the velocities of the pulsar, the Earth and the screen, respectively, $R$ is the distance from the observer to the pulsar, $r$ is the distance from the observer to the screen. The distance to the pulsar $R$ is $262 \pm 5$~pc, the transverse velocity of the visual beam $V_{psr} = 36.6 \pm 0.7$~km/s (\citeauthor{Thorsett2002}, \citeyear{Thorsett2002}). The speed of the Earth is about 30 km/s, $V_{scr}$ is about 10 km/s and it can be ignored. At a frequency of 110 MHz, we have saturated scintillation, the modulation index of intensity variations from session to session at the MP longitude, according to our observations, is 0.985.

Diffraction scintillating causes modulation of the signal intensity in both time and frequency. The time scale of the modulation in time is $t_{dif} > 200$~s (\citeauthor{Smirnova2008}, \citeyear{Smirnova2008}) and we cannot measure it. In the work \citeauthor{Bell2016} (\citeyear{Bell2016}), the value $t_{dif} = 28.8$~min was obtained at a frequency of 154 MHz. Taking the frequency dependence $t_{dif} ~ f^2 / (n-2)$ and the value of the inhomogeneity spectrum indicator $n = 3$ (\citeauthor{Smirnova2008}, \citeyear{Smirnova2008}), we get the value of $t_{dif}$ in terms of the frequency of 110 MHz =  14.7~min, i.e. significantly more time of our session.

To determine the frequency scale of scintillating at the longitudes of the MP and IP maxima, we analyzed the spectra accumulated during the session (amplitude versus frequency) in 32 channels. Since the time scale of diffraction scintillation is significantly longer than the time of a single session, time scintillation will not affect the spectrum. Such spectra for three observation sessions are shown in Fig.~\ref{fig:fig6}. It can be seen that the spectra correlate well, both in shape and scale for MP and IP. The characteristic frequency scale of scintillating $f_{dif}$ is usually determined by a shift at which the amplitude of the correlation function decreases by a factor of 2. The frequency scale varies greatly from session to session over a wide range: so for MJD 56933 $f_{dif} = 230$~kHz, and for MJD 58880 $f_{dif} = 800$~kHz. The frequency scales are the same for MP and IP.

\section{Discussion and conclusion}

Strong variations in the intensity of PSR B0950+08 emission over a wide range of time scales have been indicated in many works by (\citeauthor{Smirnova2012}, \citeyear{Smirnova2012}; \citeauthor{Singal2012}, \citeyear{Singal2012}; \citeauthor{Bhat1999}, \citeyear{Bhat1999}; \citeauthor{Bell2016}, \citeyear{Bell2016}; \citeauthor{Kuiack2020}, \citeyear{Kuiack2020}). We observe strong variations in the MP amplitude from session to session (time interval of a day or more), the modulation index is $m = 0.985$. In the work (\citeauthor{Bell2016}, \citeyear{Bell2016}) with observations at a frequency of 154 MHz, a strong variability in the flux density with a modulation index of $m = 1.3$ was indicated. As we have shown, when averaged over a long period of time, the variations in the amplitudes of the pulsar components from year to year become small, about 8\% (Table~\ref{tab:tab1}). There is a significant correlation between the MP and IP amplitude variations averaged over a time interval of 3.2 minutes and from year to year. In \citeauthor{Hankins1981} (\citeyear{Hankins1981}), analyzing the behavior of the emission intensity from pulse to pulse (over a time interval of 200 pulses) at a frequency of 430 MHz, we found a correlation between IP and MP energies with a coefficient of 0.13-0.33 for different samples. We did not find such a correlation when analyzing pulse amplitudes within a single session.

The frequency diffraction scale of B0950+08 in our observations varies greatly from session to session ($t_{dif} > 3.2$~min). Large variations in the diffraction parameters of nearby pulsars were indicated in the work \citeauthor{Bhat1999} (\citeyear{Bhat1999}) on long-term observations at 327 MHz. B0950+08 was not included in this list, however, $f_{dif}$ variations by 4-5 times are a common phenomenon. Observations of the nearby pulsar ($R = 270$~pc) B1133+16 at a frequency of 110 MHz also showed strong variations in the diffraction parameters of \cite{Popov2024}. Bell et al (\citeauthor{Bell2016}, \citeyear{Bell2016}) measured $f_{dif} = 4.1$~MHz at a frequency of 154 MHz for B0950+08. Using the dependency $f_{dif} \sim f^{-4.4}$ for the Kolmogorov spectrum, we get $f_{dif}= 930$~kHz, close to our value. The measurement of $f_{dif}$ in \citeauthor{Smirnova2008} (\citeyear{Smirnova2008}) at 112 MHz also agrees well with our data. For the first time, we measured the frequency diffraction scale at the interpulse longitude and showed that the shape of the spectrum and the frequency scale are the same for MP and IP. This fact can also serve as an argument in favor of MP and IP emission from one pole and, consequently, a small angle between the magnetic field and the axis of rotation.

The high correlation between MP and IP amplitudes from session to session at zero time shift is due to three reasons: scintillating, polarization and the ionosphere. It is difficult to separate them. The time scale of variations of 1-2 days is mainly due to refractive scintillation, since both polarization and diffraction scintillation do not correlate from day to day. In \citeauthor{Gupta1993} (\citeyear{Gupta1993}), refractive scintillating of nearby pulsars at a frequency of 75 MHz was observed, including PSR B0950+08. The authors give it $T_{ref} = 3.2 \pm 1.4$~days. Taking the dependence $T_{ref} \sim f^{-2.2}$ and extrapolating to the frequency of 110 MHz, we get $T_{ref} = 1.4$~ days or within the error range from 0.8 to 2 days, which is close to the value we received 1-2 days. For the model of scattering over an extended medium and the Kolmogorov spectrum of inhomogeneities without an internal scale, we use $m_{ref}$ \citeauthor{Rickett1984} (\citeyear{Rickett1984}) for refraction time and modulation index:

\begin{equation}
T_{ref} = \frac{0.5}{V_{ef}} \times (\frac{cR}{\pi f_{dif}})^{1/2} 
	\label{eq:5}		
\end{equation}

\begin{equation}
m_{ref} = 1.21 \times (\frac{f_{dif}}{2f})^{0.17}
	\label{eq:6}			
\end{equation}

Here $V =V_{ef}$ and for $r =R/2$ and $V_{ef} =V_{psr} + V_{obs}$ (equation 3). Taking the average value for $V_{ef} = 40$~km/s ($V_{obs}$varies significantly with time), $f_{dif} = 230$~kHz and 800 kHz, we get $T_{ref} = 4.7$~days and 2.5 days, respectively. Since the frequency diffraction scale varies greatly with time, and we do not know its average value over the entire time interval of our observations, we can say that the $T_{ref}$ predicted by the model agrees well with the value we obtained $T_{ref} = 1-2$~days. The expected value of $m_{ref} = 0.42$ and 0.52 for $f_{dif} = 230$~kHz and 800 kHz, respectively. Our modulation index $m = 1$ is significantly higher than predicted by the model. Perhaps this is due to the presence of another refractive scale of 130 days, which we also associate with refractive scintillation on large scales of inhomogeneities. The scale itself can be determined by a shift equal to $1/4$ of the distance between the CCF (df) minima in the central part in Fig. 5. It is on the order of 30 days and the spatial scale of the scattering disk is: $s =T_{ref} \times V_{ef} = 10^{13}$~cm.

\section{CONCLUSIONS}

As a result of the analysis of 9-year daily observations, the following results were obtained:

The average and relative values of the amplitudes of the components of the average profile PSR B0950+08 at a frequency of 110.3 MHz were determined when averaging profiles over the years: $A_{MP} = 23.7\pm 1.9$ Jy, $A_{IP} = 0.46\pm 0.08$ Jy, $A_m =0.24\pm 0.06$ yang, $A_{IP}/A_{MP} = 2.0 \pm 0.28\%$, $A_m/A_{MP}= 1.04 \pm 0.28\%$; the amplitude of IP may exceed the amplitude of MP in individual pulses, but this is a fairly rare event;

There are strong variations in the amplitude of the average profile from session to session on scales from one day. The average amplitude of these variations is $A = 25$~Jy, the modulation index is 1. Peak amplitudes reach 240 Jy, while $S/N = 750$;

A high correlation was found (the correlation coefficient is 0.63) between MP and IP amplitude variations both when averaging profiles over 3.2 minutes and when averaging over years. This correlation is due to refractive scintillating. No correlation of MP and IP amplitude variations was found for individual pulses within the observation session;

The frequency scale of IP diffraction scintillation was measured for the first time and it was shown that the shape of the spectrum and the scale are the same for MP and IP. There are strong variations of $f_{dif}$ from session to session (the starry day between sessions): 200 - 800 kHz;

The scale of refractive scintillation $T_{ref} = 1-2$~days was determined, variations in the amplitude of the signal with a period of about 130 days were detected. The short scale is consistent with the prediction for the extended scattering medium model with the Kolmogorov inhomogeneity spectrum. The second scale seems to be related to refractive scintillating on large inhomogeneities of the order of $10^{13}$~cm.

{ACKNOWLEDGMENTS}

The study was carried out at the expense of a grant Russian Science Foundation 22-12-00236, https://rscf.ru/project/22-12-00236/. The authors thank L.B. Potapova for help in preparing the paper.

\bsp	
\label{lastpage}

\bibliographystyle{mnras}

\end{document}